# An effective software risk prediction management analysis of data using machine learning and data mining method

Jinxin Xu
Department of Cox Business School
Southern Methodist University
Dallas, TX, USA
jensenjxx@gmail.com

Yue Wang
Center for Data Science
New York University
New York, NY, USA
yw898@nyu.edu

Ruisi Li
School of Professional Studies
Columbia University
New York, NY, USA
irisli7728@gmail.com

Ziyue Wang
Independent Researcher
New York, NY, USA
zw2013@nyu.edu

Qian Zhao
Department of Cox Business School
Southern Methodist University
Dallas, Texas, USA
scarlettzhao95@outlook.com

*Abstract*—**For one to guarantee higher-quality software development processes, risk management is essential. Furthermore, risks are those that could negatively impact an organization's operations or a project's progress. The appropriate prioritisation of software project risks is a crucial factor in ascertaining the software project's performance features and eventual success. They can be used harmoniously with the same training samples and have good complement and compatibility. We carried out in-depth tests on four benchmark datasets to confirm the efficacy of our CIA approach in closed-world and open-world scenarios, with and without defence. We also present a sequential augmentation parameter optimisation technique that captures the interdependencies of the latest deep learning state-of-the-art WF attack models. To achieve precise software risk assessment, the enhanced crow search algorithm (ECSA) is used to modify the ANFIS settings. Solutions that very slightly alter the local optimum and stay inside it are extracted using the ECSA. ANFIS variable when utilising the ANFIS technique. An experimental validation with NASA 93 dataset and 93 software project values was performed. This method's output presents a clear image of the software risk elements that are essential to achieving project performance. The results of our experiments show that, when compared to other current methods, our integrative fuzzy techniques may perform more accurately and effectively in the evaluation of software project risks.**

Manuscript received October 9, 2001. (Write the date on which you submitted your paper for review.) This work was supported in part by the U.S. Department of Commerce under Grant BS123456 (sponsor and financial support acknowledgment goes here). Paper titles should be written in uppercase and lowercase letters, not all uppercase. Avoid writing long formulas with subscripts in the title; short formulas that identify the elements are fine (e.g., "Nd–Fe–B"). Do not write "(Invited)" in the title. Full names of authors are preferred in the author field, but are not required. Put a space between authors' initials.

F. A. Author is with the National Institute of Standards and Technology, Boulder, CO 80305 USA (corresponding author to provide phone: 303-555-5555; fax: 303-555-5555; e-mail: author@ boulder.nist.gov).

S. B. Author, Jr., was with Rice University, Houston, TX 77005 USA. He is now with the Department of Physics, Colorado State University, Fort Collins, CO 80523 USA (e-mail: author@lamar.colostate.edu).

T. C. Author is with the Electrical Engineering Department, University of Colorado, Boulder, CO 80309 USA, on leave from the National Research Institute for Metals, Tsukuba, Japan (e-mail: author@nrim.go.jp).

## I. INTRODUCTION

BUSINESSES operate in a dynamic, globalised environment where innovation and market positioning drive everyday demands that are becoming more and more in number. Due to these demands, projects are created that require proper management in order to provide outcomes that adhere to the predetermined guidelines for quality, cost, and timeliness [1]. Because they are unique and have a deadline, projects are inherently uncertain. Projects are vulnerable to a range of risks, which are speculative events or situations that, should they materialise, can impact the project's goals in a favourable or unfavourable way [2].

The identification of risks and the proper handling of them are the main goals of risk management. Projects might have general or specific hazards. A particular task is allocated to the first level, and the project is related to the second. The risks are connected to the project activities and identified at the initial level [3]. Determining how people act in order to accomplish the activity's goals is a smart way to detect dangers. In this regard, models grounded in the Activity Theory [4] deal with the way in which artefacts are organised to meet the goals. Regarding that notion, Leontiev [5] Though Engeström [6] Considering Vygotsky's theory of the individual action to the activity, which is typically collective, he extended the idea of activities to a system of activities, where acts are part of a broader system of relationships with other activities.

Software development involves a lot of risk management.

Large-scale project management, according to Charrate, is essentially risk management [7]. Furthermore, according to Microsoft's research, if just 5% of the whole budget is allocated to risk management, there is a 50–70% chance that the project will be finished on schedule [8]. The pioneers of this field, initially introduced software risk management into the field of software project management in the late 1980s. Since then, software risk management has been a significant topic of study for software engineering [9, 10].

The intelligent analysis of historical data has received more attention in recent years. Software project risk can be predicted using intelligent learning techniques such as Artificial Neural Networks (ANN), Bayesian Belief Networks (BBN), and Influence Diagrams (ID). The two intrinsic drawbacks of the aforementioned techniques are that both ID and BNN include the need for past knowledge and are challenging to expand. ANN is a widely used approach for risk prediction, nevertheless, and it has shown to be successful. An ANN-based model for programme sub-module risk prediction was proposed by Khoshgoftaar. Furthermore, ANN technique outperforms other traditional methods, such as Decision Trees and Multiple Regression [11, 12], in handling complex software project risk prediction problems.

Project management has become more complex primarily as a result of organisations becoming more dynamic as a result of their ongoing efforts to find new goods and markets. Using technology as an alternative to handle this dynamicity has led to the emergence of ubiquitous computing [13] as a method to help project managers [14]. A ubiquitous computing system is contextually aware and requires little interference. These characteristics allowed the ubiquitous systems to be proactive and user-responsive [15].

The following are this paper's primary contributions:

• By combining three decision-making approaches fuzzy DEMATEL, ANFIS MCDM, and IF-TOPSIS methods—a combined fuzzy-based risk evaluation framework effectively determines and prioritises the risks associated with software projects.

• To determine the ultimate rank of the project factors, the fuzzy DEMATEL approach's priority weights for software risk criteria are first applied to the ANFIS MCDM and IF-TOPSIS techniques

• By figuring out its ideal characteristics, the ANFIS performance is enhanced using three ECSA components.

• The software specialists can address software risks that are causing the project performance to deteriorate early by using this proposed fuzzy-based risk evaluation approach.

• Six-dimensional software project risk criteria, which incorporate the NASA 93 COCOMO dataset, are experimentally validated.

## II. RELATED WORK

This paper presents a summary of several ideas regarding risk assessment and management in software initiatives, including machine learning methods and approaches that have been applied to provide more accurate assessments and categories for the danger variables and levels that can be noticed during the process of creating a project involving software, as well as conventional approaches used to detect and manage risks in software projects. Additionally, the study looks at machine learning-focused risk management research projects that are published elsewhere in order to identify the common inputs and outputs or algorithms associated with this field of study [16].

Financial risk warning is becoming a crucial component of contemporary business finance management. This research essentially proposes the optimised The BP neural network as a model for monetary warning signs guaranteeing excellent prediction accuracy. The research describes the model's operating concept and associated reasoning process, analyses its drawbacks, and suggests fixes [17].

A computational approach for risk prediction-based project failure probability reduction is proposed. The study aims to present a model that will help teams identify and track risks at various stages of a project's life cycle. This work is presented as a scholarly contribution to the inference of recommended risks for new initiatives using context histories. In the first, two teams evaluated how the prototype was used while five projects were being carried out. In the second scenario, 17 completed projects were examined in order to evaluate the Átropos model's suggestions and compare them with the hazards associated with the original projects. Átropos used 70% of each project's execution in this scenario to learn and recommend risks to be developed for the identical projects [18].

A successful Cyber Security Risk Management (CSRM) approach that takes into account asset criticality, risk type prediction, and assessment of the efficacy of current safeguards. We use a variety of techniques for the proposed unified strategy, including fuzzy-set theory for asset the point of criticality and a Comprehensive Assessment Model (CAM) to assess the effectiveness of the current regulations' classifiers that use machine learning to anticipate risk. The proposed method governs and maps relevant CSRM concepts (property, threat actor, offensive pattern, plan of action, strategy, and procedure, or TTP) with the features of the VERIS communal dataset (VCDB) in order to predict risk [19].

This article discusses the use of machine learning techniques and fuzzy set theory in risk assessment for excavation. One case study that demonstrates the application of a machine learning technique to risk assessment is the excavation at the Guangzhou metro station. The quantity of information gathered during excavation by sensors, 3S methods (RS, GIS, and GPS), and other methods improves the precision of risk assessment levels. Upon integration with a building information modelling (BIM) management platform, these protocols have the ability to oversee, regulate, and keep an eye on dynamic safety hazards. In the end, the processing and analysis of the enormous amounts of data gathered by 3S techniques and sensors has produced exciting possibilities for the creation of an integrated technology system for digging [20].

In order to promote accurate delayed project risk evaluation and forecasting using reliable data sources, pertinent delay hazards and elements first were determined, and a multivariate data set reflecting the duration of previous projects efficiency and delay-inducing risk sources was then created. Following this, a data exploratory study revealed the interrelated nature of the system, and two appropriate artificial intelligence models were selected and trained using the data set to estimate project postponement extents. Lastly, the predicted performance of each model were reviewed through crossover tests, and their results were further compared [21].

The suggested model was put into practice taking monitoring data and expert opinions into account. To increase the trustworthiness of the results, qualified verdicts were given the appropriate weight using the expert confidence index. The high-risk factors were identified using the proposed model. The suggested model was used on an excavations architecture project in Tianjin in order to assess its efficacy [22].

For large-scale building projects, a cross-analytical-machine learning model-based risk prediction system was created. On a five-point Likert scale, industry experts provided original data on 63 risk variables related to the megaproject's cost, schedule, quality, and scope. After additional statistical processing, the acquired sample produced a notably large collection of characteristics for K-means clustering based on the identification of related sub-risk components and high-risk factors. To keep the most important aspects related to cost, time, quality, and scope, descriptive analysis, the synthetic minority over-sampling technique (SMOTE), and the Wilcoxon rank-sum test were used successively [23].

This study suggests a thorough approach based on a fuzzy analytical hierarchy process (FAHP) and fuzzy Bayesian network (FBN) for the evolving risk analysis of foundational pit collapse under development. First, the findings of expert consultation, fault tree analysis, and statistical analysis of foundation pit collapse instances are used to identify the likely risk factors for underground pit disaster. Next, an FAHP and improved expertise gathering that considers an assurance index are employed to determine the odds parameters of the BN. Risk-based assessment and susceptibility analysis of foundations pit collapsing are carried out using fuzzy Bayesian inference [24].

This research proposes a comprehensive method for the changing risk analysis of foundational pit collapse under development, based on a fuzzy Bayesian network (FBN) and fuzzy analytical hierarchy process (FAHP). First, probable risk factors for basement pit collapsed are identified using the results of consulting with experts, fault tree examination, and statistical analysis of foundation pit collapse cases. Next, an FAHP and improved expertise gathering that considers an assurance index are employed to determine the odds parameters of the BN. Risk-based assessment and susceptibility analysis of foundations pit collapsing are carried out using fuzzy Bayesian inference [25].

The model incorporates the concept of entropy weight strategy, the fuzzy decision-making trial and evaluation laboratory (FDEMATEL), and the multiattributive border approximating area comparison (MABAC) approach. First, the risk variables—geology, the environment around it, monitoring, construction, and management—are thoroughly analysed. Second, the objective weight is computed using the entropy weight method, while the subjective weight is computed using the fuzzy DEMATEL methodology. The MABAC technique is then presented in order to determine the foundation pit's primary risk variables as well as the construction's level of danger [26].

This research proposes the Att-Bilstm, a hybrid deep learning model that precisely and dynamically forecasts the risk of future excavations to neighbouring buildings by integrating the self-attention mechanism with the bidirectional long and short-term memory neural network (Bi-LSTM). To improve the explanation of the Att-Bilstm model, risk evolution is examined using the Latin hypercube sampling (LHS) approach, and the Shapley Additive explanations (SHAP) analysis is used to quantify the contribution of components in risk prediction [27].

First, the intuitionistic fuzzy sets-TOPSIS framework is introduced to accurately determine the degree of rockburst severity in hydraulic tunnel. Additionally, the crucial matrix is built. Then, the weighing decisive matrix is computed. The degree of membership is ultimately ascertained at different intensities of rockburst. The risk level that matches the highest membership degree, as determined by the membership degree ranking sequence, is the ultimate evaluation level. It is concluded that the suggested approach of determining the intensity of a rockburst is workable [28, 29].

In order to precisely assess the degree of rockburst in hydraulic tunnels, we first present the intuitionistic fuzzy sets-TOPSIS model. Additionally, the crucial matrix is built. The weighed decisive matrix is then computed after that. The degree of membership at various levels of rockburst intensity ultimately determines this. The risk level that most closely matches the highest membership degree is the final evaluation level in the membership degree ranking process. The suggested approach can be utilised to calculate a rockburst's power, it is concluded [30].

The inadequacies of conventional risk management techniques [31].; the definition and application of appropriate machine learning algorithms; the management of data scarcity and uncertainty in the construction industry; the use of probabilistic graphical models to represent interdependencies between project variables and risks; the automation of risk identification and assessment processes; the ongoing evaluation and enhancement of risk management performance; and the practical implementation requirements, challenges, ethics, biases, and potential harms. Three different ML-based models were created and applied to two case studies in order to accomplish these goals. In the first case study, the industry partner for this research, Jacobs Italia SPA, had its project portfolio analyzed [32].

## III. MATERIAL AND METHODS

The accomplishment of project performance has been significantly impacted by the prioritisation of software project risks. In actuality, removing inevitable project risks has a significant impact on the evolution of the software life cycle. In order to assess the risks associated with software projects, an integrated framework for software risk evaluation is thus proposed.

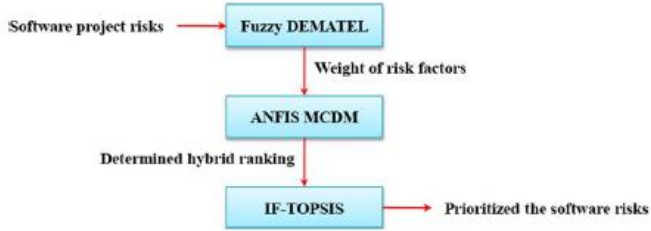

Fig. 1. Proposed framework.

The suggested framework for evaluating software risk is flow normed in Figure 1.

### A. Fuzzy DEMATEL approach

A complex measurement system finds it challenging to evaluate the exact values while making decisions. The fuzzy DEMATEL technique is a useful tool for figuring out the subjective weights given to each decision component. As a result, the fuzzy DEMATEL technique is used to calculate the ranks and weights of the risk factors as follows:
Formation of a direct relational matrix

NASA 93 COCOMO datasets are first collected from different organisations to generate a direct relational matrix. In addition, the definition of a matrix of decisions s is $(P_i, i = 1, 2, 3, \ldots, n)$. Stated differently, the n the risk criteria for software projects is shown by key weights. To ascertain the key weights, m responders (R) have submitted subjective assessments based on the language scale measurements 0–4. The threat criterion for software projects is shown by key weights. To ascertain the key weights, m respondents (R) have submitted subjective assessments based on the scale's measurements 0–4. where the matrix of direct relationships is represented by $\widehat{S}$, which is as follows:

$$R^1 \quad R^2 \quad R^3 \quad \ldots \quad R^n$$

$$
\begin{array}{l}
P_1 \\
P_2 \\
P_3 \\
\vdots \\
P_m
\end{array}
\begin{bmatrix}
\hat{s}_1^1 & \hat{s}_1^2 & \hat{s}_1^3 & \cdots & \hat{s}_1^m \\
\hat{s}_2^1 & \hat{s}_2^2 & \hat{s}_2^3 & \cdots & \hat{s}_2^m \\
\hat{s}_3^1 & \hat{s}_3^2 & \hat{s}_3^3 & \cdots & \hat{s}_3^m \\
\vdots & \vdots & \vdots & \ddots & \vdots \\
\hat{s}_m^1 & \hat{s}_m^2 & \hat{s}_m^3 & \cdots & \hat{s}_m^m
\end{bmatrix}, \quad i = 1,2,\ldots.,m; \quad j =
$$

$$1,2,\ldots.,n \qquad (1)$$

Here, m and n represent a decision matrix's dimensionalities. Where m stands for the number of criteria or factors and n for the number of responses. However, the triangular fuzzy numbers (TFNs) that are most frequently encountered are the upper (u), middle (m), and lower (l). In a similar vein, TFN has been evaluated by responders with any degree of component in the jth influential factor $\widehat{s}j = (ls\,j, ms\,j, us\,j)$.

Uncertain language variables: The aforementioned fuzzy linguistic variables can be used to address the ambiguity and uncertainty that arise throughout the human process of decision-making. Crisp scores (CFCs) are produced by converting fuzzy linguistic variables developed through direct relational matrix creation (starting matrix (S)). The initial values of the direct relational matrix (S) are found using the logistic formula. By dividing the total initial linear related matrix S by the number of people who responded (by 93), the average value for the initial matrix (S) can be found.

Equations (2) and (3) aid in the formation of the generalised relational matrix (Q). All of the matrix's diagonal members have a value of zero, and the matrix's elements must all meet the requirement that $0 \leq L_{ji} \leq 1$. The set of items (L) for the direct relational matrix (S) with n × n matrix is obtained, and then pairwise comparison is performed to ascertain the relevance of risk factors.

$$L = \frac{1}{\max\limits_{1 \leq j \leq m} \sum_{i=1}^{m} P_{ji}} \qquad (2)$$

$$Q = L \times S \qquad (3)$$

Creating the overall relationship matrix (T) Q is used to assess the total relational matrix (T) based on (4).

$$T = Q(1 - Q)^{-1} \qquad (4)$$

Calculating each risk criterion's prominence and relationship by adding up the rows and columns. The rows and columns of matrix T are denoted by the symbols $C_i$ and $R_i$, respectively. The horizontal and vertical axes are represented by the prominence and relation, respectively, according to Eqs. (5–7), and their summing is denoted as $R_i + C_i$. Since the criteria/factor $i$ depends on factor $j$, the element $t_{ij}$ denotes an indirect influence.

$$T = t_{ij}, I, j = 1, 2, \ldots, m \qquad (5)$$

$$R_i = \sum_{i \leq 1 \leq n} t_{ij} \qquad (6)$$

$$C_i = \sum_{i \leq 1 \leq n} t_{ij} \qquad (7)$$

Analysing the factors' priority weights Eq. (8) is utilised to ascertain the priority weights $w_i$ for the factors. More importantly, this estimated priority weight has been used by the ANFIS-MCDM and IFTOPSIS techniques to ascertain the effects of risk factors.

$$w_i = \frac{\sum_{j=1}^{n} R_i + C_i}{\sum_{i=1}^{n} \sum_{j=1}^{n} (R_i + C_i)} \qquad (8)$$

### B. 3.2 ANFIS MCDM approach

The decision makers will use the ANFIS MCDM technique to perform ranking, selection, and assessment processes based on the weights assigned to the problems. The systematic approach has not been applied in previous research while designing fuzzy-based systems. Fuzzy systems are created using artificial neural networks (ANNs) to overcome this constraint. They are listed in the following order:

1. Assessing the incidences of software risk to assess the likelihood of risk factors in software $(P_j,\ j = 1,2,3,\dots,n)$, a decision matrix $\hat{v}$ is created. An example of $\hat{v}$ might be this:

$$
\begin{array}{c}
\quad r^1 \quad r^2 \quad r^3 \quad \dots \quad r^n \\
\begin{array}{c}
P_1 \\ P_2 \\ P_3 \\ \vdots \\ P_n
\end{array}
\left[
\begin{array}{ccccc}
\hat{s}_1^1 & \hat{s}_1^2 & \hat{s}_1^3 & \cdots & \hat{s}_1^n \\
\hat{s}_2^1 & \hat{s}_2^2 & \hat{s}_2^3 & \cdots & \hat{s}_2^n \\
\hat{s}_3^1 & \hat{s}_3^2 & \hat{s}_3^3 & \cdots & \hat{s}_3^n \\
\vdots & \vdots & \vdots & \vdots & \vdots \\
\hat{s}_m^1 & \hat{s}_m^2 & \hat{s}_m^3 & \cdots & \hat{s}_m^n
\end{array}
\right]
\end{array}
$$

$$i = 1,2,\dots m;\quad j = 1,2,\dots,n \qquad (9)$$

The five option linguistic scales are shown in Table 2. On the other hand, the likelihood of risk severity and the chance of risk occurrences are represented by the subjective assessments. Additionally, these two risk conditions are described using language scales that include catastrophic, high, medium, low, likely, extremely likely, even, very unlikely, and substantial. Based on data regarding risk occurrences and the likelihood For every risk factor, the risk magnitude is calculated in relation to the risk severity.

Furthermore, each linguistic variable risk factor's level of risk is evaluated using the fuzzy values of the PMBOK.

1. In light of ANN's capabilities on MCDM, we therefore select an ANFIS conduct for predicting human comprehension, modelling, and development. The notion of risk degree and probability of risk occurrences are considered inputs; nevertheless, the ANFIS system determines the risk quantity by taking the risk size into account as a consequence. The high risks comprised quality factors data that was not completely developed by prior research.

modifying the ECSA optimisation strategy to enhance ANFIS functionality

Gang's fuzzy inference system (FIS) combines the concept of fuzzy logic with artificial neural networks (ANN) to streamline the learning and adaptation processes. The challenges in detecting FIS parameters are resolved by the adaptive network combined with the neurofuzzy models. It should be mentioned that the estimation of adaptive networks' overall output behaviour is done using an updated set of variables. The suggested model makes use of the Sugeno neuro-fuzzy system. The output of the Sugeno system is a linear function for each fuzzy rule. First order and zero order derivatives forms have been established through linear relations. Nevertheless, this linear relationship is displayed as a constant integer in the second order form. Lastly, the outputs obtained from each are added together using the weightage.

Using (10), two of the method's inputs are determined based on the linear relationship between them.

$$If\ (U\ is\ h_1)\ AND\ (V\ is\ g_1)\ THEN\ (X_1 = q_1 U + r_1 V + s_1) \qquad (10)$$

In this case, the numerical variables are denoted by $h$ and $g$, respectively, and inputs can be seen numerically by U and V. The membership functions serve as the basis for evaluating these variables. The application of variables/parameters $q, r,$ and $s$ yields the input and output. spatial connection.

L1: The layer 1 fuzzification technique illustrates various fuzzy sets that require numerical inputs. Using (11) the resultant output is produced from L1.

$$L_{j1} = \omega h_j(U);\quad j = 1,2;$$

$$L_{j1} = \omega h_{j-2}(V);\quad j = 3,4;$$

In this case, the fuzzy sets are represented by $g$ and $h$, and the membership functions are defined by the expressions $\omega_{h_j}(U)$ and $\omega_{g_i}(V)$. We have chosen the Gaussian kind of function of membership (shown in (12)) among others.

$$L_{j1} = \omega_{h_j}(U) = \frac{1}{1 + (\frac{(U-m_j)}{l_j})^{2k_j}} \qquad (12)$$

which shows the representation of the parameters of function as $(m_j, k_j, l_j)$

L2: The "AND" or "OR" operators are used to calculate the firing strength (output). Due to this, the output [firing strength (w)] is produced by multiplying the preceding outputs (13).

$$L_{j2} = w_j = \omega_{h_j}(U); j = 1,2 \qquad (13)$$

L3: The weights derived from L2 are normalised using Eq. (14).

$$L_{j3} = \overline{w}_j = \frac{w_j}{w_1 + w_2};\quad j = 1,2 \qquad (14)$$

L4: By figuring out whether each rule role is necessary, the model output is derived using (15).

$$L_{j4} = w_j f_j = \overline{w}_j (q_j U + r_j V + s_j);\ j = 1,2 \qquad (15)$$

The output produced from the previous layer is denoted by $w_j$.

L5: This last layer is also known as the output layer. One does summation using the other neuron's prior outputs. The de-fuzzification phase comes after the translation of fuzzy outputs to numerical outputs (16).

$$F(U,V) = \frac{w_1 f_1(U,V) + w_2 f_2(U,V)}{w_1(U,V) + w_2(U,V)} = \frac{w_1 f_1 + w_2 f_2}{w_1 + w_2}$$

$$L_{j5} = f(U,V) = \frac{\sum_j w_j f_j}{\sum_j w_j} \qquad (16)$$

In this scenario, by taking into account the supervised learning rules, the network rules can be taught. As a result, throughout the procedure, the values of the parameters $s, r, q, m, l,$ and $k$ are periodically changed. Additionally, as shown in Eq. (17)], the parameters $m, l,$ and $k$ behave as constants. and the procedure is repeated until the optimal parameter value is obtained.

$$f = \overline{w}_1(q_1(U) + r_1(V) + s_1) + \overline{w}_2(q_2 U + r_2 V + s_2)$$

$$f = (\overline{w}_1 U)q_1 + (\overline{w}_1 V)r_1 + \overline{w}_1 s_1 + (\overline{w}_2 U)q_1 + (\overline{w}_2 V)r_1 + \overline{w}_2 s_2 \qquad (17)$$

Traditionally, to find the ideal parameters for the ANFIS approach, the least squares method was utilised. A huge search space and a decreased convergence speed are obtained when applying unknown membership functions with numerous inputs.

In this article, the enhanced crow search (ECSA) optimisation strategy is used to increase the ANFIS performance.

Creation of an improved optimisation method for the crow search.

The primary drawback of CSA with a more evenly a process of balanced exploration as well as extraction is entrapment in a local optimum, as is the case with all metaheuristic algorithms. The sole major parameter that affects the improvement of the CSA's searching capability and convergence speed is awareness probability (AP). An enhanced version of CSA known as ECSA was created in order to get around this restriction.

Conversely, ECSA offered three features to enhance both regional and worldwide search functions. These are listed below: (a) a new global position updating approach; (b) a local neighbourhood searching strategy ($LN_{SS}$) is used to pick the crow that needs to follow; and (c) a dynamic AP ($D_{AP}$) strategy is used to balance intensification and diversification.

### 1) Dynamic awareness probability (DAP)

The two most important metrics are flight time (FL) as well as awareness probability (AP) crucial CSA adjustment factors. Furthermore, the CSA's AP parameter enhances the harmony between the exploration and exploitation phases. To overcome this challenge, each crow's AP value is adjusted using the DAP in accordance with the rank it obtains in each repetition.

This method calculates the fitness function for every crow. The acquired fitness value is then used in a sorting process to arrange the data in ascending order of best to worst. Lastly, each crow is given a rank, and then the sorting process is carried out in (18).

$$rank_j = j, \quad j = 1,2,\dots N_P \qquad (18)$$

According to (18), the crow with a better solution (lower fitness) receives rank 1. The crow with the weakest answer is then awarded the final rank. But each crow uses the rank it received to determine its AP value, which is among APMIN and APMAX, as seen below:

$$D_{APj}^{itr} = AP_{MIN} + (AP_{MAX} - AP_{MIN})\frac{rank_j}{N_P}$$

The rank gained for the jth solution is represented by rankj. The population size is denoted as NP, while the maximum and minimum values of AP are denoted as $AP_{MAX}$ and $AP_{MIN}$, respectively.

The parameters of A $AP_{MAX}$ and $AP_{MIN}$ have been fixed at 0.8 and 0.1 for this work.

According to the aforementioned equation, the AP value increases as the crow's rank gradually rises. In this manner, the best crow receives the top rank, and the minimum value of AP is established. On the other hand, the worst crow receives the last rank, and the AP value is set to the maximum. While a local search, or exploitation, is largely carried out in a random fashion by the best crow, exploration, or worldwide search, is most frequently carried out by the unluckiest crow.

### 2) Enhanced local search

Let s = $[U_1,\dots U_{NP}]$ and a D-dimensional parameter vector $U_j = [u_{1,1}, \dots, u_{j,D}]$ takes into account each crow $U$. In the process of updating positions, traditional CSA has determined the meal positions of randomly picked crows, resulting in a slow convergence rate. Stated otherwise, the slow convergence rate is attained if it is determined that the chosen crow is the poorest option. With each crow $i$ selecting crow $j$ as a follower, the classic CSA uses a local neighbourhood selective strategy ($LN_{SS}$) to improve its exploitation method. A crow named $U_j$ has changed its location for convenience by choosing a follower crow from a nearby little neighbourhood. Approaching the ECSA idea, every crow $U_j$ generated from the local neighbourhood takes into account a D-dimensional parameter vector $D = [D_1,\dots, D_d]$. Nonetheless, the crow $U_j$ determines the optimal location from each index $d_j$. The crow $U_j$ located in its local neighbourhood is randomly selected for each dimension, and its neighbourhood is then generated after that. It is clear that by using ECSA to update the entire dimensions of crow $U_j$ rather than modifying the same crow $U_j$, local minimal trapping and increased exploration are accomplished. The updated crow positions are displayed in Equation (20).

$$u_{i,s}^{itr+1} = u_{i,s}^{itr} + F_L^{itr} \times \left(n_{D(i),s}^{itr} - u_{i,s}^{itr}\right); \quad s = 1\dots d \qquad (20)$$

There isn't another static neighbourhood at the moment the search is being conducted. Notably, after deciding on the number of repetitions, the population's crows are randomly shuffled to create a static neighbourhood. In this way, the latest knowledge is disseminated to enhance the crow's capacity for investigation.

### 3) Improved worldwide search methodology

After migrating in an arbitrary direction and applying the knowledge that the ith crow's presence was discovered by the jth crow, convergence speed is impacted in a classical CSA. Thus, by incorporating a unique global strategy into the ECSA, the exploration capacity of the classic CSA is enhanced. The global best options that have been discovered are taken into account for the crow's position update [shown in Eq. (21)]. As a result, the crow explores and exploits the area surrounding the search area.

$$u_j^{itr+1} = Best + C1 * C2, \quad If\ rand < 0.5$$
$$u_j^{itr+1} = Best - C1 * C2, \quad otherwise \qquad (21)$$

A linearly declining variable is denoted as C1 as the iterations go, and the global optimal solution is The variable acquired from the interval [0,1] is displayed as C2, and the current position of the jth crow is indicated as $u_j$. Using Equation (22) the crow's location is updated in the neighbourhood while taking into account the global best answer.

$$C1 = 2 * exp\left(\frac{-4 * itr}{MAX_{itr}}\right)^2 \qquad (22)$$

*Why enhanced crow search*

Many scholars are interested in CSA, a new metaheuristic algorithm that is simple to use and easy to implement. Due to CSA's limited ability to maintain a better balance between the exploration and exploitation processes, a poor premature convergence rate is the outcome. This research analyses an improved version of CSA to find the best ideal settings to increase the ANFIS performance in order to overcome this problem. The population size is set at 10, approximately 20 independent runs are conducted to get all statistical findings, and a maximum of 100 iterations are used. The weighting parameter in fitness estimate has a fixed value of 0.9 for determining the classification accuracy.

ECSA is used to fix the initial value for a set of solutions. The M-dimensionality of the $N_p$ ravens in the population $P^{itr} = (itr = 0, \ldots, \max\_itr)$ is represented by the matrix below.

$$P^{itr} = \begin{bmatrix} u_{1,1}^{itr} & u_{1,2}^{itr} & \cdots & u_{1,N}^{itr} \\ \vdots & \vdots & \cdots & \vdots \\ u_{N_p,1}^{itr} & u_{N_p,2}^{itr} & \cdots & u_{N_p,N}^{itr} \end{bmatrix}$$

(23)

where the number of crows is indicated by $N_p$, the number of solutions is indicated by N, and the position of the crows is recorded in the matrix $P^{itr}$. The position of the jth agent's Dth dimension is indicated as $u_{j,D}^{itr}$ in each $iteration\ (itr)$. Using the ECSA method, the crow is first positioned towards a randomly chosen spot.

$$u_{ji} = Lw_{ji} + rand * (UP_{ji} - L_{ji}), \qquad j = 1,2,\ldots,N_p,\ i = 1,\ldots,N, \qquad (24)$$

The upper and lower boundaries utilised for the element $u_i \in U$ at ith dimension are represented as $Lw_{ji}$ and $Up_{ji}$, respectively. The equation below shows how each crow can be converted into a binary solution.

$$u_{j,D}^{itr+1} = \begin{cases} 1 & if\ S(u_{j,d}^{itr+1}) \geq \sigma \\ 0 & otherwise \end{cases} \qquad (25)$$

$\sigma \in [0,1]$ is used to represent a threshold that is represented using a random value. The continuous valued input vector $u_i$ for each generation itr with dimension D is represented as $u_{j,D}^{itr+1}$. In equation (26) the sigmoid function is shown.

$$S(u_{j,D}^{itr+1}) = \frac{1}{1 + e^{-u_{j,D}^{itr+1}}} \qquad (26)$$

The global best solution to attain maximum classifying precision in ANFIS is chosen by taking into consideration the best possible subset for every crow, using the fitness derivation below.

$$Fit = \beta . Err(S) + (1 - \beta) \qquad (27)$$

In this instance, the categorization error is represented as Err(s). The real parameter $\beta \in [1,0]$ is used to indicate the weight to be computed for the classification error rate, and $1 - \beta$ is used to denote the relevance of the best option among the options in the subset. Nevertheless, $\beta$ has a value of 0.9. The optimal answer is determined once Equations are used to modify the crow's placement (20) and (21). Based on the $D_{AP}$ value found in Eq. (19), the position is updated. Iterations of the process are carried out frequently until the ideal answer is found.

The ANFIS enters a key phase when non-normalized software project data is incorporated into the risk appraisal procedure. Therefore, to get around this problem, the ANFIS parameters must be precisely changed. An attempt has been made to use ECSA to modify the ANFIS parameters in this work. Here, the linear and Gaussian functions are taken into consideration for the input and output. The input parameters are the standard deviation, average, and output parameters consist of a constant number and an attribute coefficient. The following is an explanation of the ideal parameters for ANFIS that were determined using ECSA and the Tagaky-Sugeno fuzzy system: An application for software risk assessment is the NASA 93 COCOMO dataset. Initially, the input data from ANFIS is normalised within the interval of 0 to 1. After that, the input data is divided into the three random groups using a threefold cross-validation procedure. Two categories are used for the training process, while the remaining categories are used for the testing phase. This makes use of "genfis2" as the base fuzzy system. Sugeno type FIS is produced using genfis2 and a subtractive clustering technique. But two distinct types of input and output sets are required by this system. Using the rule of extraction approach, the antecedent and quantity of rules that define membership are assessed by the sub-clust function. The consequents of each rule are then ascertained using the least linear square estimation. The following is an explanation of the ANFIS parameter adjustment procedures.

Table 1: Pseudocode of ECSA

| Algorithm 1: |
| --- |
| Set Itr=1; |
| Initialize $N_P$ crow's position in the search space |
| Initialize the crow's memory based on the initial positions |
| Using Eq. (27) calculate the fitness of each crow |
| Based on the fitness value population (Pop) is sorted |
| While itr<max_itr |
| Using Eq. (22), set the value of C1 |
| for j=1:size(Pop) |
|     Evaluate the $D_{AP_j}^{itr}$ using Eq. (19) |
|     If rand>=$D_{AP_j}^{itr}$ |
| //Begin local search |
| Vector D is constructed using the local neighborhood of $U_j$ |
|     For i=1: dimension |
|       $u_{i,s}^{itr+1} = u_{j,s}^{itr} + F_L^{itr} \times (n_{D(i),s}^{itr} - u_{j,s}^{itr})$; $s = 1\ldots d$ |
|     End |
| Else |
| //Begin global search |
| If rand<1 |
|     $u_j^{itr+1} = Best + C1 * C2$ |
| Else |
|     $u_j^{itr+1} = Best - C1 * C2$ |
| End |
|     End for |
|     Using Eq. (25), the solutions are binarized |
|     Using Eq. (21), the positions updated are determined |
|     If found better is the obtained fitness then the crow's memory is enhanced |
|     Local neighborhood is changed |
| Itr=itr+1 |
| End while |
| Select the best optimal parameter (the global best solution) |

A preliminary fuzzy system and its interpretation parameters The inputs and outputs of the function memberships are the parameters taken into account here. These parameters are vector-stored; b) the baseline fuzzy system settings can be changed by using ECSA: The best parametric values are taken into consideration while choosing the principal parameter coefficient in this case. The parameter $i$ is assumed to have an initial value of $a_i^0$, a coefficient of $U_i$, and an ideal measure indicated as $U_i^*$ for the parameter $i$. The value of $U_i^*$ is found using equation (28).

$$a_i^* = U_i a_i^0 \qquad (28)$$

To find the coefficient $U_i$, apply ECSA. There are two important conditions that are part of the coefficient $U_i$: (1) the signs have not changed, only the size. For $\delta$, the ideal value could be $\le 1$ so that $U_i \in [10^{-\delta}, 10^{-\delta}]$. The size and signs have been changed. M may have a value $\le 10$ such that $U_i \in [-M, +M]$.

The fuzzy system is fitted with the new parameters that are obtained via ECSA. Computing the MAPE and RMSE errormetrics yields the minimal error, per Eq. (25). The best or optimal parameters found in ECSA are then stored for use in the following iteration.

Optimised settings for ANFIS Using normalised data for training and testing (the same as the step discussed earlier), a base fuzzy system known as genfis2 is utilised. However, instead of using the system settings that are set by default, the optimised parameters that were obtained following the prior stage's use of ECSA are employed. More importantly, the final MAPE and RMSE measurements are determined using the optimal parameters that were acquired. The ANFIS can accurately assess risk based on the determined ideal parameters. Every time it runs, the ECSA generates a different set of parameters (Fig. 2). As a result, the best model was used in the final version.

2. Analysing software project performance effectiveness by assessing software risk variables and assigning potential ratings. For the software risk factors, the significant weights ($w_j$) and potential scores ($f_i$) are computed. A anticipated value ($P_{out}$) for software project risks is calculated by multiplying the two derived values using Equation (29).

$$P_{out} = \sum_{j.i=1}^{n} w_i f_i \qquad (29)$$

The normalised priority weight of the jth software risk factor is displayed as $w_j$ using the Fuzzy-DEMATEL technique, and risk factors with their potential scores are indicated as $f_i$.

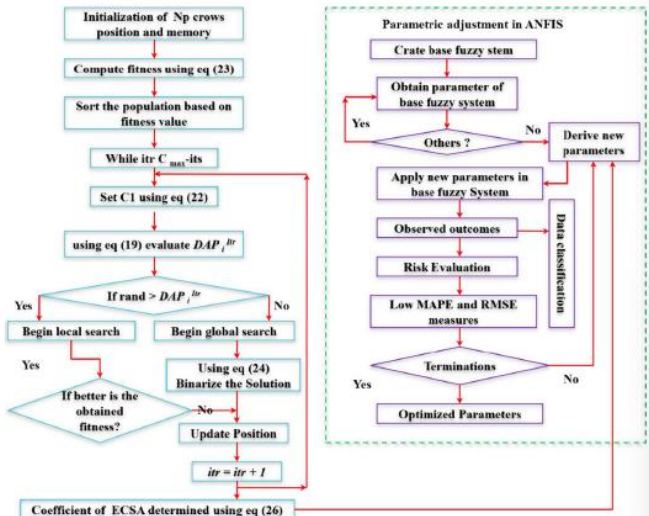

Fig. 2. Parametric optimization in ANFIS.

### C. IF-TOPSIS approach

An expansion of fuzzy TOPSIS is the intuitionistic fuzzy TOPSIS (IF-TOPSIS) method. TOPSIS that is used to calculate the dominance degree for each project choice by implementing a multi-criteria system derived from prospect theory (PT). In this work, the CFCS technique has been used to obtain a crisp score by de-fuzzing fuzzy data. In the beginning, fuzzy-DEMATEL and ANFIS MCDM were used to determine the rank and weights of the software risk variables. The final risk factor weights are then generated using the IFTOPSIS approach after first balancing the risk factor weights obtained using the fuzzy DEMATEL approach. The following is an explanation of the final weights prioritised using the IFTOPSIS technique for software project risks:

Combined weighted IF-decision matrix formation

$$S \otimes W = \{u, \mu_{Bi}(u).\mu_w(u), x_{Bi}(u) + x_w(u) - x_{Bi}(u) - x_w(u) | u \in U\}$$

and

$$\pi_{Bi_w}(u) = 1 - x_{Ai}(u) - x_w(u) - \mu_{Bi}(u).\mu_w(u) + x_{Bi}(u).x_w(u) \qquad (30)$$

The elements comprising the combined weighted IF-decision matrices are shown above as $s'_{ij} = (\mu'_{ij}, x'_{ij}, \pi'_{ij}) = (\mu_{Bi} w(u_j), x_{Bi} w(u_j), \pi_{Bi} w(u_j))$

Originate The optimum IF-positive and IF-negative solutions

Assume that benefit and cost have the respective values $I_1$ and $I_2$. The IF-positive ideal solution is $B^-$, and the IF-negative ideal solution is $B^*$. With Eqs. (32–35) in mind, the following is the extraction of (31):

$$B^* = \left( \mu_{B^* w}(u_j), x_{B^* w}(u_j) \right)$$

and

$$B^- = \left( \mu_{B^- w}(u_j), x_{B^- w}(u_j) \right), \qquad (31)$$

$$\mu_{B^* w}(u_j) = \left( \left( \max_i \mu_{B_{i w}}(u_j) | j \in I_1 \right). \left( \min_i \mu_{B_{i w}}(u_j) | j \in I_2 \right) \right) \qquad (32)$$

$$x_{B^* w}(u_j) = \left( \left( \max_i x_{B_{i w}}(u_j) | j \in I_1 \right). \left( \min_i x_{B_{i w}}(u_j) | j \in I_2 \right) \right) \qquad (33)$$

$$\mu_{B^- w}(u_j) = \left( \left( \max_i x_{B_{i w}}(u_j) | j \in I_1 \right). \left( \min_i \mu_{B_{i w}}(u_j) | j \in I_2 \right) \right) \qquad (34)$$

$$x_{B^- w}(u_j) = \left( \left( \max_i x_{B_{i w}}(u_j) | j \in I_1 \right). \left( \min_i x_{B_{i w}}(u_j) | j \in I_2 \right) \right) \qquad (35)$$

Finding the values of separation

The number of possible distance values is used to compute the spacing between options. In addition, the Euclidean, Hamming, and normalised distance metrics were applied. The separation measures $V_i^*$ and $V_i^-$ are then calculated using the positive and negative ideal solutions that pertain to each alternative. (36) and (37), which use normalised Euclidean distance, show the estimated separation measure.

$$V_i^+ = \sqrt{\frac{1}{2m} \sum_{j=1}^{m} \left[ \left( \mu_{B_{j}w}(u_j) - \mu_{B^*-w}(u_j) \right)^2 + \left( x_{B_{j}w}(u_j) - x_{B^*-w}(u_j) \right)^2 + \left( \pi_{B_{j}w}(u_j) - \pi_{B^*-w}(u_j) \right)^2 \right]} \tag{36}$$

$$V_i^- = \sqrt{\frac{1}{2m} \sum_{j=1}^{m} \left[ \left( \mu_{B_{j}w}(u_j) - \mu_{B'-w}(u_j) \right)^2 + \left( x_{B_{j}w}(u_j) - x_{B-w}(u_j) \right)^2 + \left( \pi_{B_{j}w}(u_j) - \pi_{B'-w}(u_j) \right)^2 \right]} \tag{37}$$

Relative proximity coefficient calculation

Equation (38) can be utilised to obtain the relative closeness coefficient of the intuitionistic ideal solution.

$$\zeta_i = \frac{V_i^-}{V_i^- + V_i^+} \text{ where } 0 \leq \zeta_i \leq 1 \tag{38}$$

Assessing the overall worth and ratings of prospects The $\xi_i$ values are sorted to establish the ranks of each alternative. The best options are those with higher $\xi_i$ values.

## IV.  EXPERIMENTAL RESULT

The NASA 93 COCOMO dataset, which may be accessed by the general public through the PROMISE data bank (http://promise.site.uottawa.ca/SERepository/datasets-page.html), is used in research. Table 5 presents 24 enterprise risk-related variables across five categories and 93 example cases. Thirty percent is used for assessment, while 70% percent of the dataset (a total of ninety-three project variables) is used for training. What's more, the majority of the NASA 93 dataset is in the COCOMO 81 model format. However, the COCOMO software cost model is closely linked to the risk factors that we have chosen for further study. In a similar vein, we also obtained data from the Indian software company UST, and we

| Software risk criteria's | Code | Assessment criteria of software risks |
|---|---|---|
| Schedule risk (P) | SCED (P₁) | Required development schedule |
| Product risk (Q) | RELY(Q₁) | Required software reliability |
| | DATA(Q₂) | Database size |
| | SIZE (Q₃) | Software size |
| | CPLX (Q₄) | Product complexity |
| | DOCU (Q₅) | Documentation |
| Platform risk (R) | TIME(R₁) | Execution time constraints |
| | STOR(R₂) | Main storage constraints |
| | DATA(R₃) | Database size |
| Personnel risk (S) | ACAP(S₁) | Analyst capability |
| | AEXP(S₂) | Application experience |
| | LTEX(S₃) | Language and tool set experience |
| | PCAP(S₄) | Programmer capability |
| | VEXP(S₅) | Virtual machine experience |
| | PCON(S₆) | Personnel continuity |
| Process risk (T) | TOOL(T₁) | Use of software tools |
| | SITE(T₂) | Multisite development |
| | PREC(T₃) | Precedentedness |
| | FLEX(T₄) | Development flexibility |
| | RESL(T₅) | Architecture or risk resolution |
| | TEAM(T₆) | Team cohesion |
| | PMAT(T₇) | Process maturity |
| | INCREMENTS(T₈) | Increment development |
| Reuse risk (U) | RUSE(U₁) | Required reusability |

used that data to compare how risk factors are prioritised. Seventy employees of the company participated in the validation of our suggested assessment methodology in order to determine how risk factors affect the performance of software projects.

Schedule risk, product risk, platform risk, people risk, process risk, and reuse risk are the six-dimensional risk factors for software variables that have been examined for three fuzzy-based multi-criterion decision- making techniques: fuzzy DEMATEL, ANFIS MCDM, and IF-TODIM methodologies (Fig. 7a–f)

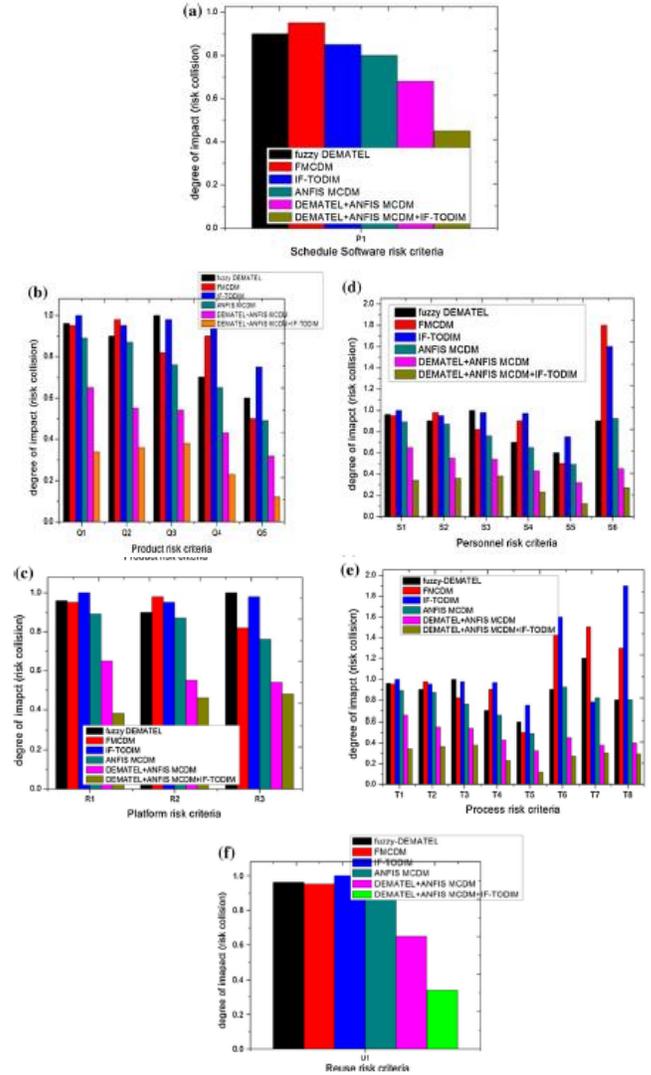

Fig. 3. The six risk factors for software risk evaluation are as follows: (a) schedule risk factor; (b) product risk factor; (c) platform risk factor; (d) personnel risk factor; (e) process risk factor; and (f) reuse risk factor.

Our recommended approach for risk assessment in an unclear setting is more accurate in its reasoning and has superior computational capacity. Nevertheless, the suggested method offers multiple ideal means for decision makers to identify risk indicators while taking human behaviour into account during the hybrid MCDM problem-solving process. Stated otherwise, our hybrid fuzzy DEMATEL+ANFIS MCDM+ IF-TODIM technique determines the triangular fuzzy numbers-based relative worth and priority of the risk components. It should be highlighted that the uncertainty issues with MCDM techniques have not been attempted to be resolved in the current research. With regard to the software project performance across four methodologies (fuzzy DEMATEL, fuzzy ANFIS MCDM, fuzzy DEMATEL, fuzzy ANFIS MCDM, and IF-TODIM), Figure 3 clearly shows the impact of each risk group. The figures show that the individual FMCDM and DEMATEL approaches, fuzzy DEMATEL, ANFIS MCDM, and IF-TODIM methodologies, as well as the hybrid (DEMATEL+ FMCDM) approach, all performed less well in software applications' risk evaluation.

The risk variables' degree of influence and linkages were evaluated using fuzzy DEMATEL, however the fuzzy-based multicriterion decision making technique (FMCDM) simply performs less well when it comes to calculating potential ratings in a fuzzy environment. Nevertheless, by examining the likelihood and seriousness of risk events and assigning a numerical risk analysis approach to the risk based on machine learning, this strategy can be further enhanced. In order to take decision makers' behaviours into account, we have finally combined the fuzzy DEMATEL+ANFIS MCDM approach with the IF-TODIM approach using triangular fuzzy numbers. The rationale for incorporating these three approaches within our suggested risk assessment framework is their ability to manage ambiguity and uncertainty in the course of decision-making, which sets them apart from other hybrid and standalone methodologies. When utilising our suggested hybrid risk assessment framework, there is a very minimal degree of risk collision because these elements are highly independent and have different ranks. This facilitates software managers' ability to better prioritise risks associated with software projects and enhance their overall performance.

## V. CONCLUSION

In the sphere of software development, the efficient prioritisation of software risks was crucial. Moreover, investigating a viable solution to this issue is a significant undertaking. Applying our recommended risk assessment framework to the combination of fuzzy DEMATEL, ANFIS MCDM, and IFTODIM methods demonstrates the efficacy of software risk factors. The study's findings indicate that the main variables influencing software performance improvement are the risk factors. Furthermore, a good software management plan should not make software projects perform worse by removing the hazardous variables; our suggested risk assessment methodology more than satisfies this need by giving the dangerous factors priority. In order to further refine this strategy and classify the decision-making problems, we want to incorporate other software risk criteria in the future. Furthermore, we plan to include our suggested framework into increasingly complex software process evaluation assignments and examine increasingly advanced machine learning methodologies.

## ACKNOWLEDGMENT

None.